# Effects of the interaction between the magnetic moments of the proton and electron on the energy states of hydrogen atom.

**Voicu Dolocan**
*Faculty of Physics, University of Bucharest, Bucharest, Romania*

We make a comparison between the energy levels of the hydrogen atom, calculated by using standard methods, and that by using a modified Coulomb potential due to the interaction between the magnetic moments of the proton and electron. In this later method we use to ways. One is that in which we solve the Schroedinger equation with the modified Coulomb potential and some constraint conditions. The other is that in which we expand the modified Coulomb potential in Taylor series. The obtained results show that the first way gives a better agreement with experimental data.

## 1. State of the Art

The hydrogen problem is viewed as a positive, motionless,central point charge, the proton, and an orbiting negative point charge, the electron, bound by the attraction of opposite charges. The attraction results in a Coulomb potential which is given by

$$V(r) = -\frac{e^2}{4\pi\varepsilon_o}\frac{1}{r} \qquad (1)$$

This potential when is plugged into the generic radial Schroedinger equation gives

$$H_o R_{nl} = E_n R_{nl}$$
$$H_o = -\frac{\hbar^2}{2\mathrm{m}}\frac{d^2}{dr^2} + \left[-\frac{e^2}{4\pi\varepsilon_o}\frac{1}{r} + \frac{\hbar^2}{2\mathrm{m}}\frac{l(l+1)}{r^2}\right] \qquad (2)$$

where $l$ is the orbital quantum number. By solving this equation one obtains the electron energy levels for the hydrogen atom

$$E_n = -\frac{e^2}{8\pi\varepsilon_o a_o n^2} \qquad (3)$$

where

$$a_o = \frac{4\pi\varepsilon_o\hbar^2}{me^2} = 0.529177\times10^{-10}\ \mathrm{m}$$

is known as the Bohr radius.. There is the $n^2$-fold degeneracy of states with the same principal quantum number, or $2n^2$ –fold once the spin degrees of freedom is included.

The degeneracy is lifted by corrections that arise due to perturbation Hamiltonians giving the fine and hyperfine structure of the hydrogen atom. The fine structure of the hydrogen atom corrections consists of three perturbation Hamiltonians. The first one is the relativistic correction

$$H_{rc}=-\frac{\left(p^2\right)^2}{8\mathrm{m}^3 c^2} \qquad (4)$$

which arises from the expansion of the relativistic kinetic energy

$$\sqrt{m^2 c^4+p^2 c^2}=mc^2+\frac{p^2}{2\mathrm{m}}-\frac{\left(p^2\right)^2}{8\mathrm{m}^3 c^2}+\ldots$$

The second is the spin-orbit coupling

$$H_{LS}=g\frac{1}{4\mathrm{m}^2 c^2}\frac{1}{r}\left(\frac{dV}{dr}\right)(L.S) \qquad (5)$$

where V is the central potential (1) for the hydrogen atom. The third and the last one is the so-called Darwin term

$$H_D=\frac{\hbar^2}{8\mathrm{m}^2 c^2}\frac{e^2}{\varepsilon_o}\delta(r) \qquad (6)$$

The first-order energy shifts due to these perturbations are simple expectation values with respect to the unperturbed states. The expectation value of the perturbation Hamiltonians for the state $|njlm_j>$ is

$$\langle njlm_j|H_{rc}+H_{LS}+H_D|njlm_j\rangle=\frac{\alpha^4 mc^2}{8\mathrm{n}^4}\left(3-\frac{8\mathrm{n}}{2\mathrm{j}+1}\right) \qquad (7)$$

This correction depends only on *j*, so that occurs degeneracy between $2s_{1/2}$ and $2p_{1/2}$, between $3s_{1/2}$ and $3p_{1/2}$, between $3p_{3/2}$ and $3d_{3/2}$, etc. The Dirac equation predicts ( as well as the Bohr-Sommerfeld quantization condition)

$$E_{nj}=mc^2\left\{1+\left(\frac{\alpha}{n-j-\frac{1}{2}+\sqrt{\left(j+\frac{1}{2}\right)^2-\alpha^2}}\right)^2\right\}^{-1/2}$$

and once expanded up $O(\alpha^4)$, we find

$$E_{n,j}=mc^2\left\{1-\frac{\alpha^2}{2n^2}+\frac{\alpha^4(6j+3-8n)}{8(2j+1)n^4}+\ldots\right\} \qquad (8)$$

which agree with Eq.(7). The degeneracy between $2s_{1/2}$ and $2p_{1/2}$, etc., is lifted by the Lamb shift. This is an effect understand only in the Quantum Electrodynamics[1], which takes the zero-point fluctuations of photons, as well as the polarization of the "vacuum" filled with negative energy electrons into account. It is well known that the electron posses a magnetic moment $\boldsymbol{\mu} = (e/2mc)(\mathbf{L}+2\mathbf{S})$. However, like the electron, the proton has spin angular momentum with $s_p = ½$, and associated with this angular momentum is an intrinsic dipole moment

$$\mu_p=\gamma_p\frac{e}{Mc}S_p \qquad (9)$$

where M is the proton mass and $\gamma_p$ is a numerical factor, known experimentally to be $\gamma_p$=2.7928. The proton dipole moment will interact with both the spin dipole moment of the electron and the orbital dipole moment of the electron, and there are two new contributions to the Hamiltonian, the nuclear spin-orbit interaction and the spin-spin interaction. These perturbation Hamiltonians give the hyperfine splittings of the energy levels. The nuclear spin-orbit Hamiltonian is

$$H_{pLS}=\frac{\gamma_p e^2}{mMc^2 r^3}L.S_p \qquad (10)$$

The spin-spin Hamiltonian is

$$H_{ss}=\frac{\gamma_p e^2}{mMc^2}\left\{\frac{1}{r^3}\left[3(s_p.r)(s_e.r)-(s_p.s_e)\right]+\frac{8\pi}{3}(s_p s_e)\delta(r)\right\} \qquad (11)$$

Consider tha case $l = 0$, because the hyperfine splitting of the hydrogen atom ground state is of the most interest. Since in this case the electron has no orbital angular momentum ($\mathbf{L}$ = 0) there is no nuclear spin –orbit effect. It can be shown that because the wave function has spherical symmetry only the delta function term contributes from the spin-spin Hamiltonian. First order perturbation theory yields the hyperfine energy shift for $l$ = 0

$$E_{hf}=\frac{m}{M}\alpha^4 mc^2\frac{4\gamma_p}{3n^3}\left[f(f+1)-\frac{3}{2}\right] \qquad (12)$$

The quantum number $f$ has possible values $f = j + ½, j – ½$, since the proton spin is ½ and $j$ is the total quantum number of the electron. It is observed that the hyperfine splittings are smaller than fine structure by a factor of m/M. For the ground state of the hydrogen atom ($n$ = 1), the energy separation between the states $f$ = 1 and $f$ = 0 is

$$E_{hf}(f=1)-E_{hf}(f=0)=5.9\times10^{-6}\ \text{eV}$$

The photon corresponding to the transition between these two states has wavelength $\lambda$ = 21.1 cm.

Another point of view is presented in [2], where we have used in Schroedinger equation a modified Coulomb potential

$$V(r)=\frac{2\hbar c}{144\pi r}\left[1+0.65\cos\left(\frac{a}{r}\right)\right]$$
$$a=\frac{\pi e^2}{mc^2}\left[\frac{2\text{m}}{M}\gamma_p s_p-m_l-2\text{m}_s\right] \qquad (13)$$

$s_p = \pm \frac{1}{2}$ is the proton spin quantum number, $m_s = \pm 1/2$ is the electron spin quantum number and $m_l$ is the magnetic quantum number. By using some constraint conditions it was found the hydrogen energy levels

$$E_n=-\frac{\hbar^2}{2\text{m}}\left[1+0.65\cos\left(\frac{a}{r_o}\right)\right]^2\frac{\left[\sqrt{n^2+\frac{4mca}{144\pi\hbar}\times1.3\sin\left(\frac{a}{r_o}\right)}-n\right]^2}{\left[1.3\text{a}\sin\left(\frac{a}{r_o}\right)\right]^2} \qquad (14)$$

where the quantity $(a/r_o)$ is that which assure the minimum value of $E_n$. It is obtained that at the minimum, $sin(a/r_o) \sim 10^{-6}$ , so that our assumption for deduction of Eq. (14) may be accepted. ( We note that in the deduction of Eq. (14) it is assumed that in preceding equation, $sin(a/r)$ and $cos(a/r)$ are practically independent of $r$, that is they have the constant values corresponding to the minimum of the energy (14)). At the minimum value of $E_n$, stationary state, $r_o$ is the electron radius $\sim n^2 a_o$. The important result is that the Eq. (14) lifts the degeneracy between $2s_{1/2}$ and $2p_{1/2}$, etc, as it is shown in Table II, Ref.[2].

In this paper, we continue the idea of the paper [2] and we will see what happens when we expand in series the *cosine* term in Eq. (13).

**2. Series expansion of the modified Coulomb potential.**

As we have seen in [2], $a/r \sim 10^{-6}$, so that $cos(a/r)$ may be expanded in series up to $O(a^4/r^4)$

$$\cos(a/r) = 1 - \frac{a^2}{2r^2} + \frac{1}{24}\frac{a^4}{r^4} - \ldots$$

So

$$V(r) = -\frac{3.3\hbar c}{144\pi r} + \frac{0.65\hbar c}{144\pi}\frac{a^2}{r^3} - \frac{0.65\hbar c}{144\pi}\frac{a^4}{12r^5} \qquad (15)$$

The first term from the right hand side is the well known Coulomb potential, $\alpha\hbar c/r$, and the next two terms are perturbations. The Hamiltonian of the system is

$$H = H_o + H'_1 + H'_2$$
$$H_o = -\frac{\hbar^2}{2m}\left(\frac{d^2}{dr^2} + \frac{2(l+1)}{r}\frac{d}{dr}\right) - \frac{3.3\hbar c}{144\pi r} \quad = -\frac{0.65\hbar c a^4}{144\pi \times 12 r^5} ¿ \qquad (16)$$
$$H'_1 = \frac{0.65\hbar c a^2}{144\pi r^3}\ ; \qquad \{H_2$$

The first order energy shifts due to these perturbations are simple expectation values with respect to the unperturbed states. The only tricky aspect is that the unperturbed states are degenerate. Therefore,

$$E'_1 = \langle nl|H'_1|nl\rangle = \frac{0.65\hbar c a^2}{144\pi}\langle\frac{1}{r^3}\rangle$$
$$E'_2 = \langle nl|H'_2|nl\rangle = -\frac{0.65\hbar c a^4}{144\pi \times 12}\langle\frac{1}{r^5}\rangle \qquad (17)$$
$$\langle r^k\rangle = \int_0^\infty r^{2+k}|R_{nl}(r)|^2\,dr$$

The values of $E'_1$ and $E'_2$ for hydrogen atom levels are presented in Table I.

Table I

The corrections to the ground state of the hydrogen atom, $E'_1$ and $E'_2$

| State | $E'_{1,\times 10^{-7}}$eV | $E'_{2,\times 10^{-11}}$eV | $E''_1, \times 10^{-7}$eV |
|---|---|---|---|
| $1S_{1/2,1/2}$ | 8.033137 | 1.27 | 341.502 |
| $1S_{1/2,-1/2}$ | 8.128975 | 1.29 | 345.576 |
| $2S_{1/2,1/2}$ | 0.988561 | 0.16 | 42.7477 |
| $2S_{1/2,-1/2}$ | 0.994589 | 0.16 | 43.0083 |
| $2P_{1/2}$ | $5.218389\times10^{-8}$ | $1.02\times10^{-15}$ | $5.218389\times10^{-8}$ |
| $2P_{3/2,1/2}$ | 0.091481 | $3.0649\times10^{-3}$ | 0.081481 |
| $2P_{3/2,-1/2}$ | 0.091540 | $3.0836\times10^{-3}$ | 0.0911540 |
| $3S_{1/2,1/2}$ | 0.291999 | 0.05 | 12.6043 |
| $3S_{1/2,-1/2}$ | 0.293779 | 0.06 | 12.6812 |
| $3P_{1/2}$ | $1.567226\times10^{-8}$ | $1.1719\times10^{-17}$ | $1.567226\times10^{-8}$ |
| $3P_{3/2,1/2}$ | 0.027105 | $3.529\times10^{-5}$ | 0.027105 |
| $3P_{3/2,-1/2}$ | 0.027188 | $3.55\times10^{-5}$ | 0.027188 |
| $3D_{3/2,1/2}$ | 0.000252 | $6.36577\times10^{-9}$ | 0.000252 |
| $3D_{3/2,-1/2}$ | 0.000254 | $6.44363\times10^{-9}$ | 0.000254 |
| $3D_{5/2,1/2}$ | 0.002274 | $5.17723\times10^{-7}$ | 0.002274 |
| $3D_{5/2,-1/2}$ | 0.002279 | $5.37136\times10^{-7}$ | 0.002279 |
| $4S_{1/2,1/2}$ | 0.116135 | 0.02 | 5.32647 |
| $4S_{1/2,-1/2}$ | 0.116843 | 0.02 | 5.35894 |
| $4P_{1/2}$ | $3.306868\times10^{-8}$ | $2.6262\times10^{-17}$ | $3.306868\times10^{-7}$ |
| $4P_{3/2,1/2}$ | 0.057176 | $7.8339\times10^{-5}$ | 0.057176 |
| $4D_{3/2,1/2}$ | 0.000571 | $3.3090\times10^{-9}$ | 0.000571 |
| $4D_{3/2,-1/2}$ | 0.000574 | $3.4409\times10^{-9}$ | 0.000574 |
| $4D_{5/2,1/2}$ | 0.005148 | $2.7643\times10^{-7}$ | 0.005148 |
| $4D_{5/2,-1/2}$ | 0.005159 | $2.7755\times10^{=7}$ | 0.005159 |
| $4F_{5/2,1/2}$ | 0.000818 | $2.5975\times10^{-9}$ | 0.000818 |
| $4F_{5/2,-1/2}$ | 0.000821 | $2.6133\times10^{-9}$ | 0.000821 |
| $4F_{7/2,1/2}$ | 0.003273 | $4.1295\times10^{-8}$ | 0.003273 |
| $4F_{7/2,-1/2}$ | 0.003287 | $4.1750\times10^{-8}$ | 0.003287 |

We have used the values of *a* given in Table II.

Table II

Values of the parameter *a*

| State | $m_l$ | $m_s$ | $s_p$ | $a$, $\times10^{-15}$ m |
|---|---|---|---|---|
| $nS_{1/2,1/2}$ | 0 | ½ | ½ | 8.83910580397 |
| $nS_{1/2,-1/2}$ | 0 | ½ | -1/2 | 8.86601104077 |
| $nP_{1/2}$ | 1 | -1/2 | ±1/2 | 0.01345262138974 |
| $nP_{3/2,1/2}$ | 1 | ½ | ½ | 17.69166422937 |
| $nP_{3/2,-1/2}$ | 1 | ½ | -1/2 | 17.718569447215 |
| $nD_{3/2,1/2}$ | 2 | -1/2 | ½ | 8.83910580389 |
| $nD_{3/2,-1/2}$ | 2 | -1/2 | -1/2 | 8.86601104677 |
| $nD_{5/2,1/2}$ | 2 | ½ | ½ | 26.54422265475 |
| $nD_{5/2,-1/2}$ | 2 | ½ | -1/2 | 26.57112789753 |
| $nF_{5/2,1/2}$ | 3 | -1/2 | ½ | 17.69166422937 |
| $nF_{5/2,-1/2}$ | 3 | -1/2 | -1/2 | 17.718569447215 |
| $nF_{7/2,1/2}$ | 3 | ½ | ½ | 35.39678392921 |
| $nF_{7/2,-1/2}$ | 3 | ½ | -1/2 | 35.42368632291 |

The radial wave function used are given in Table III.

Table III. Radial wave functions.

| $n$ | $l$ | $R_{nl}$ |
|---|---|---|
| 1 | 0 | $2\left(\frac{1}{a_u}\right)^{3/2} e^{-r/a_o}$ |
| 2 | 0 | $\left(\frac{1}{2a_o}\right)^{3/2}\left(2-r/a_o\right)e^{-r/2a_o}$ |
| 2 | 1 | $-r/2a_o$<br>$\left(\frac{1}{2a_o}\right)^{3/2}\frac{1}{\sqrt{3}}\frac{r}{a_o}$ *ealignl* ¿¿¿ |
| 3 | 0 | $2\left(\frac{1}{3a_o}\right)^{3/2}\left(1-\frac{2}{3}\frac{r}{a_o}+\frac{2}{27}\left(\frac{r}{a_o}\right)^2\right)e^{-r/3a_o}$ |
| 3 | 1 | $\left(\frac{1}{3a_o}\right)^{3/2}\frac{4\sqrt{2}}{3}\left(1-\frac{1}{6}\frac{r}{a_o}\right)\frac{r}{3a_o}e^{-r/3a_o}$ |
| 3 | 2 | $\left(\frac{1}{3a_o}\right)^{3/2}\frac{2\sqrt{2}}{27\sqrt{5}}\left(\frac{r}{a_o}\right)^2 e^{-r/3a_o}$ |
| 4 | 0 | $\frac{1}{96}\left(\frac{1}{a_o}\right)^{3/2}\left(24-36\frac{r}{2a_o}+12\left(\frac{r}{2a_o}\right)^2-\left(\frac{r}{2a_o}\right)^3\right)e^{-r/4a_o}$ |
| 4 | 1 | $\frac{1}{32\sqrt{15}}\left(\frac{1}{a_o}\right)^{3/2}\left(20-10\frac{r}{2a_o}+\left(\frac{r}{2a_o}\right)^2\right)\frac{r}{2a_o}e^{-r/4a_o}$ |
| 4 | 2 | $\frac{1}{96\sqrt{5}}\left(\frac{1}{a_o}\right)^{3/2}\left(6-\frac{r}{2a_o}\right)\left(\frac{r}{2a_o}\right)^2 e^{-r/4a_o}$ |
| 4 | 3 | $\frac{1}{96\sqrt{35}}\left(\frac{1}{a_o}\right)^{3/2}\left(\frac{r}{2a_o}\right)^3 e^{-r/4a_o}$ |

From Table I results that the energy of perturbation is much smaller than those obtained from the standard theory or the theory which use the modified Coulomb potential and constraint conditions [2]. However, it is observed that the energy of the $nP_{1/2}$ state is lower than the energies of the other energy states (Lamb shift). An important observation is that for the evaluation of $<1/r^k>$ we have cut the lower limit of the integral to the value $\lambda_c = \hbar/mc$ (Compton wavelength), then when the integrand ~ $1/r^k$, $k \geq 1$. For example, the integral for $2S_{1/2,1/2}$ was calculated from $\lambda_c$ to $\infty$, while the integral for $2P_{1/2}$ was calculated from 0 to $\infty$. As it result from Table I,

$$\Delta E_1 = E_1^{'}(2P_{1/2}) - E_1^{'}(2S_{1/2,-1/2}) \approx 10^{-7} \text{ eV}$$

The experimental value of the Lamb shift [3] is ~ $4.38\times10^{-6}$eV In order to obtain $\Delta E_1' = 4.38\times10^{-6}$eV, we must take the lower limit corresponding to $l = 0$, at $10^{-77}$. The results are denoted by $E''_1$ and are presented in the last column of the Table I. But, if we use this lower limit to calculate $R''_2$ , one obtains enormous values. Therefore the application of the series expansion (15) appears to be inadequate.

**3. Conclusions**

We have compared the results for the hydrogen energy levels obtained by standard methods (Bohr-Sommerfeld and Dirac) with the results obtained by using a modified Coulomb potential [2]. Further, we have calculated these levels by using a series expansion of the modified Coulomb potential. We have concluded that by solving the Schroedinger equation with a modified Coulomb potential and some constraint conditions[2] one obtains the nondegenerate energy levels given by Eq. (14) which are comparable with experimental results (Table II, Ref.[2]). As it result from Section 2 of this paper, by using the series expansion (15), the degeneracy is likewise lifted, but appear difficulties due to divergence of some integrals.